\documentclass[]{aa}
\usepackage{graphicx}
\usepackage{amssymb,amsmath}
\usepackage[comma]{natbib}
\newcommand{\mdens}{{\rm g~cm^{-3}}}

\newcommand{\msun}{{\rm M}_\odot}

\usepackage{color}

\def\drom{{\rm d}}

\begin{document}
\title{Formation scenarios and mass-radius relation for neutron stars}
\author{J.L. Zdunik \and P. Haensel}
 \institute{N. Copernicus Astronomical Center PAS,
Bartycka 18, PL-00-716 Warszawa, Poland\\
{\tt jlz@camk@edu.pl, haensel@camk.edu.pl}} \offprints{J.L. Zdunik}
%
\abstract{}{Neutron star crust,  formed via accretion of matter from a companion in
a low-mass X-ray binary (LMXB), has an equation of state (EOS)  stiffer than that
of catalyzed matter. At a given neutron star mass, $M$, the radius of a  star with
an accreted crust is therefore larger, by $\Delta R(M)$,  than for usually
considered star built of catalyzed matter.} {Using a compressible liquid drop model
of nuclei, we calculate, within
 the one-component plasma  approximation,
 the EOSs corresponding to different nuclear compositions of ashes of
 X-ray bursts in LMXB.
These  EOSs are then applied for studying the effect of different formation
scenarios on the neutron-star mass-radius relation} {Assuming  the SLy EOS for
neutron star's liquid core, derived by Douchin \& Haensel (2001),  we find that at
$M=1.4~{\rm M}_\odot$ the star with accreted crust has a radius more than 100 m
larger that for the crust of catalyzed matter. Using smallness of the crust mass
compared to $M$, we derive a formula that relates $\Delta R(M)$  to the difference
in the crust EOS. This very precise formula gives also analytic dependence of
$\Delta R$ on $M$ and $R$ of the reference star built of catalyzed matter. The
formula is valid for any EOS of the liquid core. Rotation of neutron star makes
$\Delta R(M)$ larger. We derive an approximate but very precise formula that gives
difference in equatorial radii, $\Delta R_{\rm eq}(M)$, as a function of stellar
rotation frequency. }{} 
\titlerunning{Formation and $R(M)$ for neutron stars}
\authorrunning{J.L. Zdunik, P. Haensel}
\maketitle
\section{Introduction}
\label{sect:intro} Establishing via observations the mass versus radius relation
for neutron stars is crucial for determination of the equation of state (EOS) of
 dense matter. It is expected that finding even a few points
 of the $M(R)$ curve could  severely limit the range of considered
 theoretical models of  matter in a liquid neutron star core.
 General method of determining the EOS at supranuclear densities
 (i.e. densities larger than the normal nuclear density $\rho_0=2.7\times
 10^{14}~\mdens$) from the $M(R)$ curve was  developed by \citet{Lindblom1992}.
 A  simplified method of determining a supranuclear segment of the EOS from
 three  measured points of the $M(R)$ has been recently described in
 \citep{Ozel2010_PRL}. In these studies, the EOS of neutron star crust
 is considered as well established and fixed. This assumption is
 valid, within relatively small uncertainties, provided the crust matter
 is in full thermodynamical equilibrium (catalyzed matter, corresponding
 at $T=0$ to the ground state of the matter).
 Such a condition is not fulfilled for the crust formed by accretion
 of matter onto a neutron star surface, from a companion
 in a low-mass X-ray binary (accreted crust, \citet{HZ1990a_heat,HZ1990b_EOS}).
 The EOS of accreted crust is stiffer than that of the ground-state one
 \citep{HZ1990b_EOS}. Therefore, at the same stellar mass, $M$,
 the radius of the neutron star with an accreted crust is larger by
 $\Delta R(M)$ than that of the star with crust built of catalyzed matter
 (catalyzed crust).  We expect that the millisecond pulsars,
 spun-up by accretion in LMXBs
  \citep{BhattaHeuvel1991} have accreted crusts, different from the
  catalyzed crusts of pulsars born in supernova explosions.

 In the present note we study the effect of the formation scenario on the
 radius-mass relation for neutron stars. Formation scenarios and corresponding equations of
state of the crust are presented in Sect.\;\ref{sect:scenarios-EOSs}. In
Sect.\;\ref{sect:R-acc-cat} we calculate  $\Delta R(M)$ for non-rotating stars,
assuming  different crust formation scenarios. Numerical results for $\Delta R(M)$,
at fixed  EOS of the liquid core,  are presented
  in Sect.\;\ref{sect:exact-R-acc-cat}. In Sect.\;\ref{sect:chi-method}
  we derive an approximate, but very precise, general formula that
relates the difference in radii  to the difference in the EOSs of the crust. The
dependence of $\Delta R(M)$ on the EOS of the core enters via a dependence on the
mass and radius of the reference star with catalyzed crust. Rotation of neutron
star increases $\Delta R(M)$, and in Sect.\;\ref{sect:rotation} we study $\Delta
R(M)$ (defined as the difference between the equatorial radii) for rotating neutron
stars. Section\;\ref{sect:summary} contains a summary of our results,
 discussion,  and conclusion.

Calculations that led to the present note were inspired by questions
from the audience during a talk by one of the authors (P.H.) at the CompStar
Workshop "Neutron star physics and nuclear physics", held at GANIL,
 Caen, France, February 14-16,  2010.
\section{Scenarios and equations of state}
\label{sect:scenarios-EOSs}
\subsection{Catalyzed crust}
\label{sect:catalyzed}
We assume that neutron star was born in a core-collapse supernova. Initially, the
outer layers of the star are hot and fluid. Their composition corresponds to
nuclear equilibrium, because at $T\gtrsim 10^{10}$\;K all reactions are
sufficiently rapid.
 The crust solidifies in the process of cooling of a newly born neutron star.
 One assumes, that while cooling and solidifying, the outer layers keep the nuclear
  equilibrium, and after reaching the strong degeneracy limit the state of the
  EOS of the crust is well approximated by that of the
  {\it cold catalyzed matter}, corresponding (in the $T=0$ limit)
   to the ground of the matter (for a detailed
discussion, see, e.g., \citet{NSbook2007}). This  EOS of the crust will be denoted
as ${\rm EOS}[{\rm cat}]$.
\subsection{Accreted crusts}
\label{sect:accreted}
We assume that neutron star has been remaining $\sim 10^{8}-10^9$~yr in a LMXB. Its
crust was formed via accretion of matter onto neutron star surface  from a
companion in the binary. Typical accretion rates in LMXBs are $\sim
10^{-10}-10^{-9}~{\rm M}_\odot~{\rm yr}^{-1}$. Therefore, as a result of the
accretion stage, the original catalyzed crust formed in scenario described in
Sect.\;\ref{sect:catalyzed} (of typical mass $\sim 10^{-2}~\msun$) has been fully
replaced by the accreted one.

{\it Accretion and X-ray bursts.}  Many LMXBs are sites of type I X-ray bursts
(hereafter: X-ray bursts) which are thermonuclear flashes in the surface layers of
accreting neutron star \citep{WoosleyTaam1976,MaraschiCavaliere1977,Joss1977}. In
the simplest model of the X-ray bursts, accreted matter, composed mostly of $^1{\rm
H}$, cumulates on the star surface and undergoes compression due to the weight of
the continuously  accreting  matter. The accreted layer is also heated by the
plasma hitting the star surface and transforming its kinetic energy into heat.
Compressed and heated hydrogen layer burns steadily, in its bottom shell with
$\rho\sim 10^5~\mdens$, into $^4{\rm He}$. The helium produced in the burning of
hydrogen is accumulating in a growing He layer beneath the H-burning shell. After
some recurrence time (typically $\sim$ hours), the helium burning is ignited at the
bottom of the He layer, typically at $\rho\sim 10^6~\mdens$. The He burning starts
in a strongly degenerate plasma (temperature $\sim 10^8$K and $\rho\sim
10^6~\mdens$). Therefore, He burning is thermally unstable and proceeds initially
in an explosive detonation mode, with local temperature exceeding $10^9$~K, and
burns the overlaying He and H layers into elements with mass number $A\sim 50-100$
(see next paragraph). Finally, the thermonuclear explosion develops into a
thermonuclear flash of the surface layer, observable as an X-ray burst. After $\sim
$ minutes H-He envelope has been transformed into a layer of nuclear ashes. The
energy released in thermonuclear burning has been radiated in an X-ray burst.
Continuing accretion leads again to the cumulation of the H-He fuel for the next
X-ray burst and the cycle accretion-burst repeats in a quasi-periodic way every few
hours.

{\it Ashes of X-ray burst.} The composition of ashes from thermonuclear burning of
an accreted H-He layer deserves a more detailed  discussion (see, e.g.,
\citet{BeardWiescher2003}). Early calculations indicated that thermonuclear
explosive burning   produced mostly $^{56}{\rm Ni}$ which then converted into
$^{56}{\rm Fe}$ by the electron captures \citep{Taam1982,AyasliJoss1982}. This
picture had  to be revised in the light of more recent  simulations of nuclear
evolution during cooling following the temperature peak of $\sim 2\times 10^9$\;K.
These simulations have shown that after a few minutes after the initial temperature
peak,  nuclear ashes contain a mixture of nuclei with $A\sim 50-100$
\citep{Schatz2001}.

{\it Reactions in accreting crust.} During accretion, the crust is a site of
exothermic reactions in a plasma which is far from the catalyzed state. A layer of
ashes of X-ray bursts is compressed under the weight of cumulating overlaying
accreted layers. The ashes of density greater than $10^5~\mdens$ are a plasma
composed of nuclei immersed in an electron gas.  The temperature in the deeper
layer of thermonuclear ashes (a few $10^8$~K at the depth of a few meters) is too
low for the thermonuclear fusion to proceed: the nuclei have $Z\sim 30-50$, and the
fusion reactions are blocked by the Coulomb barriers. Therefore, the only nuclear
processes are those induced by compression of matter. These processes are: electron
captures, and neutron emissions and absorptions. Compression of ashes results in
the increasing density of their layer (and its increasing depth below the stellar
surface), and leads to the electron captures on nuclei and the neutronization of
the matter. To be specific, let us consider a neutron star with $M=1.4\;\msun$
(Fig.\; 39 in \citet{ChamelHaensel2008-rev}). For densities exceeding $\sim 5\times
10^{11}~\mdens$ (neutron drip point in accreted crust,  at the depth $\sim 300$\;m)
electron captures are followed by the neutron emission from nuclei. Therefore,
apart from being immersed in electron gas, nuclei become immersed in a neutron gas.
At the density greater than $10^{12}~\mdens$ (depth greater than $\sim 350$\;m) the
value of $Z$ becomes so low,   and the energy of the quantum zero-point motion of
nuclei around the lattice sites so high, that the pycnonuclear fusion  (see, e.g.,
\citet{ST1983}) of neighboring nuclei might be possible. This would lead to a
further neutronization of the considered layer of accreted matter. With increasing
depth and density, the element of matter under consideration becomes more and more
neutronized, and the fraction of free neutrons outside nuclei in the total number
of nucleons increases. At the density $10^{14}~\mdens$ (depth $\sim 1\;$km) the
crust matter element dissolves into a homogeneous $n-p-e$ plasma, containing only a
few percent of protons.

{\it Nuclear composition of accreted crusts.} The composition of a fully accreted
crust (all the crust, including its bottom layer,  being obtained by compression of
an initial shell with $\rho=10^{8}~\mdens$) is calculated, assuming a simple model
of one-component plasma, as in \citep{HZ2003_A_i}. Two initial values of the mass
number of nuclei in the  X-ray bursts ashes are assumed, $A_{\rm i}=56,\; 106$.
Thermal effects are neglected. Extensive numerical simulations of the nuclear
evolution of multi-component ashes, assuming large reaction network, and taking
into account temperature effects, were carried out by \citep{Gupta2007_multi_plas}.
However, the latter  calculations were restricted to densities less than
$10^{11}~\mdens$, where the EOS is not significantly different from  ${\rm
EOS}[{\rm cat}]$\;, albeit the nuclear composition is very different from the
catalyzed matter one.

A model of the  EOS of a fully accreted crust will be denoted as ${\rm EOS}[{\rm
acc}.{\cal A}]$, where ${\cal A}$ refers to the assumptions underlying the model:
composition of X-ray bursts ashes, types of reactions included in the model, etc.
\subsection{Equations of state}
\label{sect:EOS-acc-cat}
\begin{figure}[h]
\resizebox{3.6in}{!} {\includegraphics[angle=0,clip]{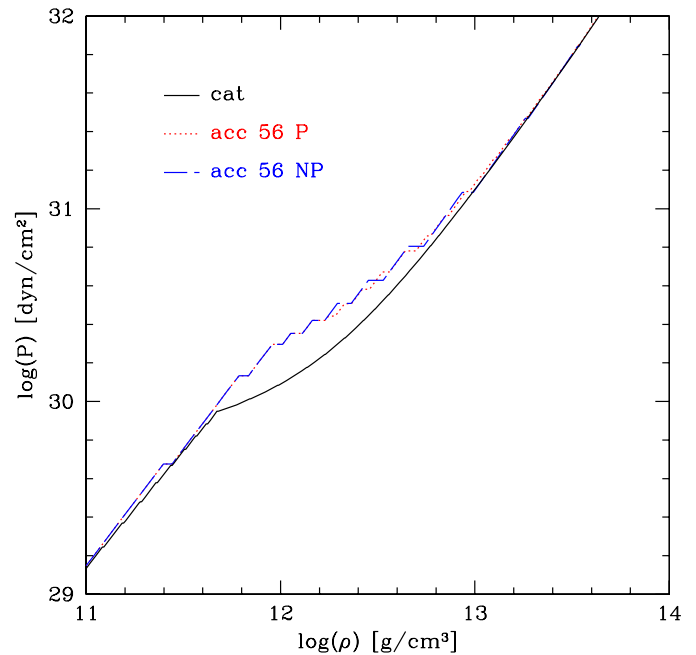}} \caption{
(color online)EOS for the crusts calculated using the Mackie-Baym model of nuclei
and neutron matter. Explanation of the labels: {\bf cat} - cold catalyzed matter;
{\bf acc.56.P} - initial X-ray burst ashes $A_{\rm i}=56$, pycnonuclear fusion
switched on; {\bf acc.56.NP} - initial X-ray burst ashes $A_{\rm i}=56$,
pycnonuclear fusion switched off till $10^{13}~\mdens$. For further details see
Sect.\;\ref{sect:EOS-acc-cat}. } \label{fig:EOSs-crust}
\end{figure}

To disentangle the effect of formation scenario from the EOS of the crust, one has
to compare  the EOSs calculated not only for the same nuclear Hamiltonian, but also
for the same model of nuclei and of nuclear matter in nuclei and  neutron matter
matter outside nuclei. In this respect analysis of \citep{HZ1990b_EOS} was not
correct, because in that paper the EOSs of accreted and catalyzed crusts were based
on different nuclear models.  In the present paper we use consistently the
compressible liquid drop model of nuclei of \citep{MackieBaym1977}: this model was
used in all previous calculations of accreted crusts. We calculated several EOSs of
accreted crust corresponding to different $A_{\rm i}$ of X-ray bursts ashes and for
different models of pycnonuclear reaction rates. These EOSs are plotted in
Fig.\;\ref{fig:EOSs-crust}, where we show also ${\rm EOS}[{\rm cat}]$ for the same
model of the nucleon component  (nuclei and neutron gas) of dense matter. For
$\rho<10^{11}~\mdens$ all EOSs are very similar, and therefore we display EOS plots
only for $\rho>10^{11}~\mdens$.

{\it Stiffness of the EOS.} Significant differences in this property of the EOS
exist in the inner crust, from the neutron drip point on, and up to
$10^{13}~\mdens$. In this density range ${\rm EOS}[{\rm acc}]$ are significantly
stiffer than ${\rm EOS}[{\rm cat}]$. The difference starts already at the neutron
drip point, which in the accreted crust is found at  a higher density. The
softening that follows after the neutron drip is much stronger in ${\rm EOS}[{\rm
cat}]$ than in ${\rm EOS}[{\rm acc}]$. Then, for $\rho>10^{13}~\mdens$  ${\rm
EOS}[{\rm acc}]$ converge (from above)  to ${\rm EOS}[{\rm cat}]$, because at such
high density the nuclei play a lesser r{\^o}le in the EOS and the crust pressure is
mostly determined by the neutron gas outside nuclei.

{\it Jumps and smoothness.} ${\rm EOS[acc]}$ shows numerous pronounced density
jumps at constant pressure, to be contrasted with smooth curve of EOS[cat].
Actually, both features are artifacts of the models used. For example, ${\rm
EOS}[{\rm cat}]$ is completely smooth, because we used a compressible liquid drop
model without shell correction for neutrons and protons, and treated  neutron and
proton numbers within the Wigner-Seitz cells as continuous variables, in which we
minimized the enthalpy per nucleon at a given pressure. For ${\rm EOS}[{\rm acc}]$
we used one-component plasma approximation, with integer neutron and proton numbers
 in the Wigner-Seitz cells. As we neglected thermal effects, we got
 density jump at each threshold for electron capture. Had we used a multicomponent
 plasma model and  included thermal effects and large reaction network, the jumps in
 the ${\rm EOS}[{\rm acc}]$ would be  smoothed (see \citet{Gupta2007_multi_plas}).

{\it Pycnonuclear fusion.} The  process of pycnonuclear fusion (see, e.g.,
\citep{ST1983}) may proceed after electron captures followed by neutron emission -
 a reaction chain which results in decreasing $Z$ and $A$. In our simulations,
based on the one-component plasma model, pycnonuclear fusion proceeds as soon as
the time $\tau_{\rm pyc}$ (inverse of the pycnonuclear fusion rate)  is smaller
than the timescale $\tau_{\rm acc}$ needed for a matter element to pass, due to
accretion, across the (quasistationary) crust shell with specific $(A,Z)$. This
time can be estimated as  $\tau_{\rm acc}=M_{\rm shell}(A,Z)/{\dot M}$. Typical
values of the mass of shells in the inner accreted crust are $M_{\rm shell}\sim
10^{-5}\;\msun$. At accretion rates ${\dot M}\sim 10^{-10}-10^{-9}\;\msun~{\rm
yr}^{-1}$ formulae used to calculate the fusion rates in
\citep{HZ1990a_heat,HZ2003_A_i} lead to pycnonuclear fusions proceeding at $\rho\ga
10^{12}~\mdens$. However, theoretical evaluation of $\tau_{\rm pyc}$ is plagued by
huge uncertainties (see \citet{Yakovlev2006}). It is not certain whether
pycnonuclear fusions do indeed occur below $10^{13}~\mdens$. If they do not, then
${\rm EOS}[{\rm acc}]$ is quite well represented by models with pycnonuclear fusion
switched off. The two extremes - pycnonuclear fusions starting at $10^{12}~\mdens$,
and pycnonuclear fusion shifted to densities above $10^{13}~\mdens$, correspond to
 curves  ${\rm EOS}[{\rm acc}.{A_{\rm i}}.{\rm P}]$ and
 ${\rm EOS}[{\rm acc}.{A_{\rm i}}.{\rm NP}]$ in Fig.\;\ref{fig:EOSs-crust}.
\vskip 2mm
 {\it Bottom crust layer with $\rho>10^{14}~\mdens$.} Our  plots of EOS in
Fig.\;\ref{fig:EOSs-crust} are restricted to  $\rho<10^{14}~\mdens$. For the
densities above $10^{14}~\mdens$ and up to the  crust bottom density $\rho_{\rm
b}$, the precision of our  models lowers significantly compared to the precision of
our EOS at lower densities. Fortunately, the contribution of the bottom layer of
the crust ($10^{14}~\mdens<\rho<\rho_{\rm b}$) to the difference $R_{\rm acc} -
R_{\rm cat}$ is negligible and therefore the uncertainties in the crust EOS at
highest densities do not affect our main results. This favorable situation will be
justified  in Sect. 3.2.

\section{Effect of the change of the crust EOS on the star radius}
\label{sect:R-acc-cat}
\subsection{Accreted vs. catalyzed crust: numerical results for non-rotating
models} \label{sect:exact-R-acc-cat}
 We matched  EOSs of the crust described
in the preceding section to several EOS of the liquid core. We checked that the
difference $\Delta R(M)= R_{\rm acc}(M) - R_{\rm cat}(M)$ does not depend on the
details of matching of the crust-core interface.  On the other hand, $\Delta R(M)$
can be shown to depend on the EOS of the core via the factor $(R_{\rm cat}/r_{\rm
g}-1)R_{\rm cat}$, where $r_{\rm g} = 2GM/c^2= 2.96~M/\msun~$km is the
gravitational (Schwarzschild) radius. These two properties  will be derived using
the equations of hydrostatic equilibrium  in Sect.\;\ref{sect:chi-method}.

\begin{figure}[h]
\resizebox{3.6in}{!} {\includegraphics[clip]{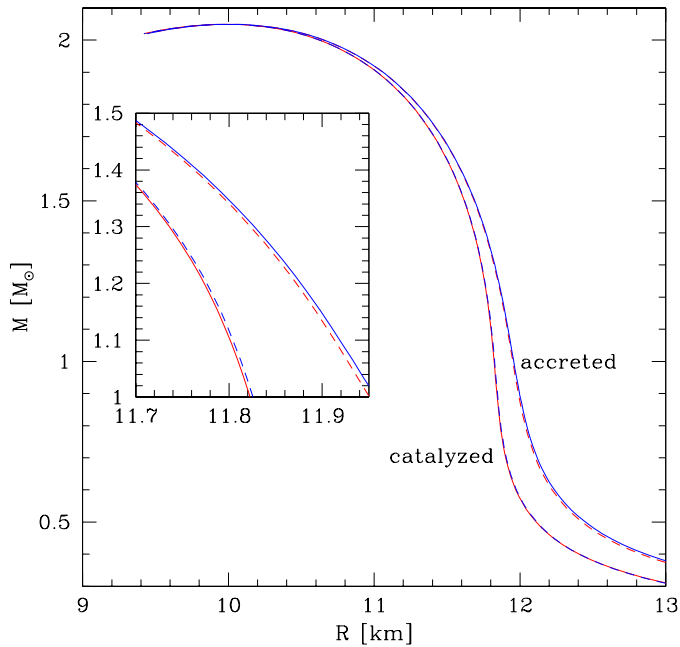}} \caption{(Color online)
Solid lines give gravitational mass $M$ versus radius $R$ for non-rotating neutron
stars with accreted [acc.56.P] and catalyzed crust. Dashed red lines correspond to
the approximation of the crust thickness by Eq.\;(\ref{eq:R.acc.cat}). Zoomed
region illustrates a very high accuracy of this approximation. For the explanation
of the procedure of getting dashed lines see the last fragment of Sect.\;3.2. }
 \label{fig:mr-chi}
\end{figure}

In Fig.\;\ref{fig:mr-chi} we show the $M(R)$ relation for neutron stars with
catalyzed and accreted crust (EOS[acc.56.P]). For the liquid core we use the SLy
EOS of \citep{DH2001}. For $M=1.4\;\msun$ we get $\Delta R\approx 100$\;m. The
value of $\Delta R$  grows to 200 m if  $M$ decreases  to $1\;\msun$. It decreases
to 40\;m  for $1.8\;\msun$.

\subsection{An analytic approximation for $\Delta R$}
\label{sect:chi-method}
Let us  consider hydrostatic equilibrium of a non-rotating neutron star. Let the
(circumferential) radius of the star be $R$ and its (gravitational) mass be $M$.
The pressure at the bottom of the crust is $P_{\rm b}$. Let the mass of the crust
be $M_{\rm cr}$. We assume that $M_{\rm cr}/M\ll 1$, but we account for the radial
extension of the crust and the radial dependence of the pressure and of the density
within the crust, following the method formulated in \citep{Zdunik2008-AB}. We use
the fact that within the crust $P/c^2\ll \rho$ and $4\pi r^3 P/Mc^2\ll 1$. We
define a dimensionless function of pressure within the crust ($0\le P\le P_{\rm
b}$)
\begin{equation}
\chi(P)=\int_0^{P}
 \frac{{\rm d} P^\prime}{\rho(P^\prime)c^2}~.
\label{eq:chi-P}
\end{equation}
The function $\chi(P)$ is determined solely by the EOS of the crust.

Using  the Tolman-Oppenheimer-Volkoff equation of hydrostatic equilibrium
(\citet{ST1983}), and neglecting within the crust  $P/c^2$ compared to $\rho$ and
to $M/(4\pi r^3)$, we can go over in Eq.\;(\ref{eq:chi-P}) to the radial coordinate
$r$, getting
\begin{equation}
\chi[P(r)]=\frac{1}{2}{\rm ln}
\left[
\frac{1-r_{\rm g}/R}{1-r_{\rm g}/r}
\right]~,
\label{eq:chi-r}
\end{equation}
where $r_{\rm g}\equiv 2GM/c^2$.

\begin{figure}[h]
\resizebox{3.6in}{!} {\includegraphics[angle=0,clip]{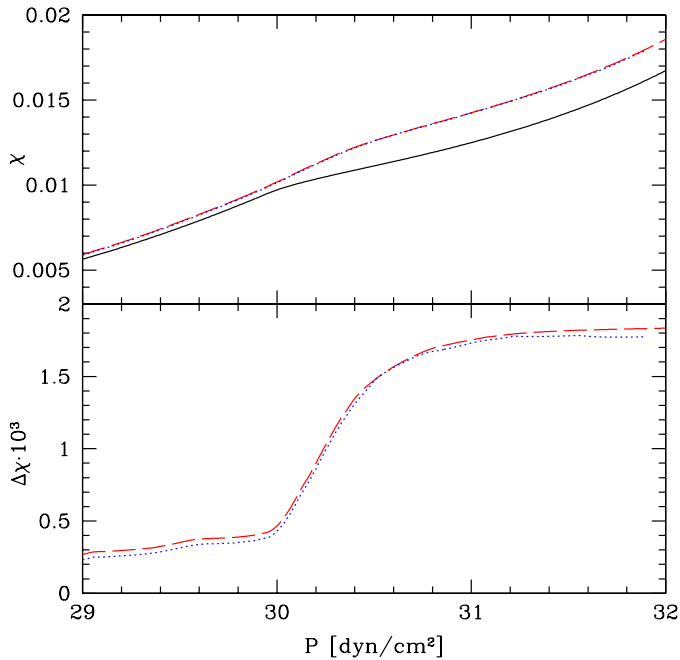}} \caption{(color
online) Upper panel: functions  $\chi(P)$ for EOS[acc.56.P] (dashed, red),
EOS[acc.106.NP] (dotted, blue)  and EOS[cat] (solid, black). Lower panel: function
$\Delta\chi(P)$ (Eq. \ref{eq:Delta.EOS}) for
 EOS[acc.56.P] (dashed, red) and  EOS[acc.106.NP] (dotted, blue)}
\label{fig:dchip}
\end{figure}

The dimensionless function $\chi\ll 1$ can be treated as a small parameter in
systematic expansions of the crust thickness. This function increases monotonously
 with $P$, from zero
to $\sim 10^{-2}$  (upper panel of Fig.\;\ref{fig:dchip}).
 Therefore, an expression for the crust layer thickness, from
the surface to the pressure at the layer bottom $P$, $\mathfrak{t}(P)\equiv
R-r(P)$, obtained from Eq.\;(\ref{eq:chi-r}) in the linear approximation in $\chi$,
is expected to be very precise,
\begin{equation}
\mathfrak{t}(P)=2\chi(P)\left(\frac{R}{r_{\rm g}}-1\right)R~.
\label{eq:d.P}
\end{equation}
Our aim is to evaluate the change in the radius of neutron star of mass $M$ and
radius $R$, when  ${\rm EOS}[{\rm cat}]$  is replaced by ${\rm EOS}[{\rm acc}]$.
These EOSs differ for pressures below $P_1=10^{32}~{\rm erg~cm^{-3}}$
(Sect.\;\ref{sect:EOS-acc-cat} ). To calculate $R_{\rm acc}-R_{\rm cat}$ it is
therefore sufficient to know  the difference in values of $\chi$ at $P_1$. We
introduce a function $\Delta\chi(P)$, measuring the difference between two EOSs of
the crust for pressures from zero to  $P$,
\begin{equation}
\Delta\chi(P)\equiv \chi_{\rm acc}(P)-\chi_{\rm cat}(P)~.
\label{eq:Delta.EOS}
\end{equation}

In Fig.\;\ref{fig:dchip} we show functions $\chi(P)$ and $\Delta\chi(P)$ for
several  EOS of accreted crust. In spite of the jumps in density at constant
pressure, characteristic of EOS of accreted crusts (Fig.\;\ref{fig:EOSs-crust}),
both $\chi$ and $\Delta\chi$ are smooth functions of $P$. This is due to the
integration over $P^\prime<P$ in the definition of  $\chi(P)$. As seen in Fig.\;
\ref{fig:dchip}, dependence of $\chi$ and $\Delta\chi$ on the particular scenario
underlying ${\rm EOS}[{\rm acc}]$  is very weak. Additionally,
 $\Delta\chi$ is nearly constant for $P>2\times 10^{31}~{\rm erg~cm^{-3}}$,
 because EOS[cat] and EOS[acc] converge at high pressures (see Fig.\;\ref{fig:EOSs-crust}).
 Therefore, $\Delta\chi(P_1)$ is a very good approximation for $\Delta\chi(P_{\rm b})$.

Our final formula, obtained using Eqs.\;(\ref{eq:d.P}),(\ref{eq:Delta.EOS}),
combined with approximations explained above, is
\begin{equation}
R_{\rm acc}-R_{\rm cat}\simeq 2\Delta[\chi]\cdot
 \left({R/r_{\rm g}}-1\right){R}~,
\label{eq:R.acc.cat}
\end{equation}
where  $\Delta[\chi]=\Delta\chi(P_1)$ and  $R=R_{\rm cat}$.

The precision of the approximation (\ref{eq:R.acc.cat}) is illustrated in
Fig.\;\ref{fig:mr-chi}. The dashed line for the accreted crust was obtained from
exact $R_{\rm cat}(M)$ using Eq.\;(\ref{eq:R.acc.cat}) with $R=R_{\rm cat}$. On the
other hand, the dashed line for catalyzed crust was obtained from exact $R_{\rm
acc}(M)$ using Eq.\;(\ref{eq:R.acc.cat}) with $R=R_{\rm acc}$.

Up to this point, we neglected contribution of elastic strain to the stress tensor
within the crust, and used the perfect liquid approximation in the equations of
hydrostatic equilibrium. Recent numerical simulations indicate, that on the
timescales of years, and longer (which are of astrophysical interest), the breaking
stress of the ion crystal of the crust is $\sim 10^{-3}$ of the crust pressure
\citep{ChugunovHorowitz2010}. This is much smaller than the difference between
EOS[acc] and EOS[cat]. Therefore, as long as we restrict ourselves to timescale
longer than a few years, contribution of elastic strain to $\Delta R(M)$ can be
neglected.

\section{Combined effects of accretion and rotation on $\Delta R(M)$}
\label{sect:rotation}
Millisecond pulsars are thought to be recycled old ("dead") pulsars, spun-up by
accretion of matter from their companion in LMXBs
\citep{AlparCheng1982,BhattaHeuvel1991}. This scenario is corroborated by discovery
of rapid X-ray pulsations with frequencies up to $619$~Hz, in more than a dozen
LMXBs. The most rapid millisecond pulsar (isolated one) rotates at 716 Hz. To reach
such high rotation frequencies, neutron star had to accrete some five times the
mass of the crust of a $1.4\;\msun$ neutron star. Therefore, one has to conclude
that millisecond pulsars have fully accreted crusts.

\begin{figure}[h]
\resizebox{3.6in}{!} {\includegraphics[clip]{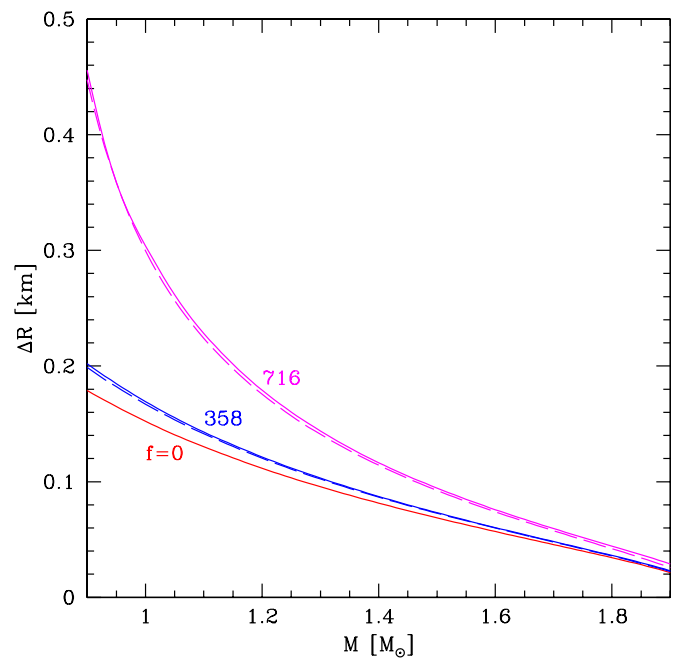}}
\caption{
Dependence of $\Delta R(M)= R_{\rm acc}(M)- R_{\rm cat}(M)$ on the frequency
of rotation of neutron star, $f$. Equations of state as in
Fig.\;\ref{fig:mr-chi}. Upper curve - $f=716\;$Hz. Middle curve -
$f=358\;$Hz. Bottom curve - non-rotating.
}
\label{fig:R-acc-cat-rot}
\end{figure}

In the case of rotating stars we will define $R_{\rm acc}-R_{\rm cat}$ as  the
difference in circumferential equatorial radii.  Rotation increases the difference
 $R_{\rm acc}-R_{\rm cat}$ compared to the static case.  A rough
Newtonian argument relies on the proportionality of the centrifugal force to the
distance from the rotation axis. Therefore, centrifugal force that acts against
gravity is largest at the equator. In order to get precise value of the effect of
rotation on $R_{\rm acc}-R_{\rm cat}$, we performed 2-D simulations of stationary
configurations of rigidly rotating neutron stars with catalyzed and accreted
crusts. The calculations have been performed using the {\tt rotstar} code from the
LORENE library ({\tt http://www.lorene.obspm.fr}. Our results, obtained for
$f=716\;$Hz (maximum frequency measured for a pulsar) and for a half of this
frequency, are shown in Fig.\;\ref{fig:R-acc-cat-rot}.

Consider hydrostatic equilibrium of the crust of rotating star in the equatorial
plane. In the Newtonian approximation, the ratio of centrifugal force to the
gravitational pull at the equator is, neglecting rotational deformation,
$\Omega^2R^3/GM$, where $\Omega$ is angular frequency of rigid rotation. Notice
that this expression is exact in the quadratic approximation in $\Omega$, because
increase of $R$ due to to rotation introduces higher powers of $\Omega$. We propose
modeling centrifugal-force effect within crust by modifying
Tolman-Oppenheimer-Volkoff equation in the equatorial plane in following way:
\begin{equation}
 \frac{\drom P}{\drom r}=-\frac{GM\rho}{r^2 }\left(1-\frac{2GM}{rc^2}\right)^{-1}
\left(1-\alpha \Omega^2 R^3/GM\right) \label{cfforce}
\end{equation}
where $M$ is the total mass of the star and $\alpha$ is a numerical coefficient to
be determined by fitting the exact results of 2-D calculations. The validity of
Eq.\;(\ref{cfforce}) relies on the smallness of the rotational flattening of the
liquid core compared to that of crust. We included terms quadratic in $\Omega$ and
we used standard approximation valid for a low-mass crust and applied in the
preceding section. Equation (\ref{cfforce}) could be solved explicitly assuming
constant $M$ and taking into account the changes of $r$ throughout  the crust, as
in the case of non-rotating star in \citep{Zdunik2002}. However,  to be consistent
with the approach leading to formula (\ref{eq:d.P}),  we prefer to  use a solution
method based on the smallness of $\chi$. Rotation effect enters via a constant
factor $\left(1-\alpha \Omega^2 R^3/GM\right)$. This results in a simple relation
between $\Omega=0$ and $\Omega>0$ difference in radii:
\begin{eqnarray}
\Delta R_{\rm eq} (\Omega)&=&R_{\rm eq, acc}(\Omega)-R_{\rm eq,cat}(\Omega)\nonumber\\
&=&\frac{2\Delta[\chi]\cdot
 \left({R/r_{\rm g}}-1\right){R}}{1-\alpha\Omega^2R^3/GM}=\frac{\Delta R (\Omega=0)}{1-\alpha\Omega^2R^3/GM}~.
 \label{eq:dromega}
\end{eqnarray}
The dimensionless parameter $\alpha$ has been determined  numerically by fitting
formula (\ref{eq:dromega}) to the exact 2-D results obtained in General Relativity
using LORENE numerical library. We found that $\alpha\approx 4/3$. As one sees in
Fig.\;\ref{fig:R-acc-cat-rot}, our approximate formula for  $\Delta R_{\rm eq}
(\Omega)$ works extremely well. The actual $\Omega$-dependence of $\Delta R_{\rm
eq}$ is stronger than quadratic because $\Omega^2$ appears in the denominator in
Eq.\;(\ref{eq:dromega}).

\section{Summary  and conclusion}
\label{sect:summary}
In the present note we studied the effect of the formation scenario on the
mass-radius relation for neutron stars. For a given $M$, a star with an accreted
crust has larger radius, by $\Delta R(M)$, than a star built of catalyzed matter
formed in stellar core collapse. We derived an approximate but very precise formula
for $\Delta R(M)$, valid for slowly rotating neutron stars. $\Delta R(M)$
factorizes into a prefactor depending solely on the EOS of neutron star crusts
formed in different scenarios and a simple function of $M$ for a given EOS of the
core.  We studied the dependence of the difference between  the equatorial radii on
the angular rotation frequency, $\Delta R_{\rm eq}(\Omega)$. We derived  an
approximate formula for $\Delta R_{\rm eq}(\Omega)$, that reproduces with high
precision $\Delta R_{\rm eq}$ even for neutron stars rotating at $716$~Hz, highest
rotation frequency measured for a radio pulsar.

We found that an accreted crust makes the radius of a $2~\msun-1~\msun$ star some
$50-200$~m larger than in the standard catalyzed matter case. Highest hopes of a
simultaneous measurement of a neutron star  $M$ and $R$ are, in this decade,
associated with high resolution X-ray spectroscopy
\citep{Arzoumian2009,Paerels2009}. Unfortunately, expected uncertainty in
determining $R(M)$ is  $\pm 5\%$. It  significantly exceeds  effects of formation
scenarios, calculated in the present note.

\acknowledgements This work was partially supported by the Polish MNiSW  grant no.N
N203 512838. This work was also supported in part by CompStar, a Research
Networking Programme of the European Science Foundation and the LEA Astro-PF.



\begin{thebibliography}{}{

\bibitem[Alpar et al.(1982)]{AlparCheng1982}
Alpar, M.A., et al., 
1982, Nature, 728

\bibitem[Arzoumian et al. (2009)]{Arzoumian2009}
Arzoumian, Z., et al., 2009, X-ray Timing of Neutron Stars, Astrophysical Probes of
Extreme Physics, arXiv:0904.3264[astro-ph.HE], A White Paper submitted to
 "Astro2010 Decadal Survey of Astronomy and Astrophysics"

\bibitem[Ayasli \& Joss (1982)]{AyasliJoss1982}
Ayasli, S., Joss, P.C., 1982,
ApJ 256, 267


\bibitem[Beard \& Wiescher (2003)]{BeardWiescher2003}
Beard, M., Wiescher, M., 2003, Revista Mex. de Fisica 49, supplemento 4, 139

\bibitem[Bhattacharya  \& van den Heuvel (1991)]{BhattaHeuvel1991}
Bhattacharya, D., van den Heuvel, E.P.J., 1991,
Phys. Rep. 203, 1


\bibitem[Chamel \& Haensel (2008)]{ChamelHaensel2008-rev}
Chamel, N., Haensel, P., 2008, Physics of Neutron Star Crusts,  Living Rev.
Relativity 11, (2008), 10. http://www.livingreviews.org/lrr-2008-10


\bibitem[Chugunov  \& Horowitz (2010)]{ChugunovHorowitz2010}
Chugunov, A.I., Horowitz, C.J., 2010, MNRAS 407, L54

\bibitem[Douchin \& Haensel(2001)]{DH2001}
Douchin, F., Haensel, P., 2001 A\& A, 380, 151


\bibitem[Gupta et al. (2007)]{Gupta2007_multi_plas}
Gupta, S., Brown, E.F., Schatz, H., M{\"o}ller, P., Kratz, K.-L.,
2007, ApJ, 662, 1188


\bibitem[G{\"u}ver et al. (2010)]{GuverOzel2010_4U_1608}
G{\"u}ver, T.,  {\"O}zel, F., Cabrera-Lavers, A., Wroblewski, P.,
2010, ApJ, 712, 964


\bibitem[Haensel \& Zdunik(1990a)]{HZ1990a_heat}
Haensel, P., Zdunik, J.L., 1990a, A\& A, 227, 431


\bibitem[Haensel \& Zdunik(1990b)]{HZ1990b_EOS}
Haensel, P., Zdunik, J.L., 1990b, A\& A, 229, 117


\bibitem[Haensel \& Zdunik(2003)]{HZ2003_A_i}
Haensel, P., Zdunik, J.L., 2003, A\& A, 404, L33


\bibitem[Haensel et al. (2007)]{NSbook2007}
Haensel, P., Potekhin, A.Y., Yakovlev, D.G. 2007, Neutron Stars 1. Equation of
State and Structure (New York, Springer)


\bibitem[Haensel \& Zdunik(2008)]{HZ2008_heat}
Haensel, P., Zdunik, J.L., 2008, A\& A, 480, 459



\bibitem[Joss (1977)]{Joss1977}
Joss, P.C. 1977, Nature 270, 310


\bibitem[Lindblom (1992)]{Lindblom1992}
Lindblom, L. 1992, ApJ 398, 569


\bibitem[Mackie \& Baym (1977)]{MackieBaym1977}
Mackie, F.D., \& Baym, G. 1977,
Nucl. Phys. A, 285, 332

\bibitem[Maraschi \& Cavaliere  (1977)]{MaraschiCavaliere1977}
Maraschi, L., Cavaliere, A. 1977, in Highlights of Astronomy, vol. 4, ed. E.A.
Mueller (Dordrecht: Reidel) Part I, p. 127


\bibitem[{\"O}zel (2006)]{Ozel2006_EXO_0748}
{\"O}zel, F., 2006, Nature, 441, 1115
ApJ, 693, 1775

\bibitem[{\"O}zel et al. (2009)]{Ozel2009_EXO_1745}
{\"O}zel, F., G{\"u}ver, T., Psaltis, D., 2009,
ApJ, 693, 1775



\bibitem[{\"O}zel et al. (2010)]{Ozel2010_PRL}
{\"O}zel, F., Baym, G., G{\"u}ver, T.2010, Phys. Rev. D, 82, 101301


\bibitem[Paerels et al. (2009)]{Paerels2009}
Paerels, F., et al., 2009, The Behavior of Matter Under Extreme Conditions,
 arXiv:0904.0435[astro-ph.HE], A White Paper submitted to
 "Astro2010 Decadal Survey of Astronomy and Astrophysics"



\bibitem[Schatz et al. (2001)]{Schatz2001}
Schatz, H., et al. 2001,
Phys. Rev. Lett. 86, 3471


%
\bibitem[Shapiro  \& Teukolsky(1983)]{ST1983}
Shapiro S.L., Teukolsky, S.A., 1983, Black Holes, White Dwarfs, and Neutron Stars:
The Physics of Compact Objects, Wiley, New York


\bibitem[Taam (1982)]{Taam1982}
Taam, R.E.,  1982,
ApJ 258, 761

\bibitem[Woosley \& Taam (1976)]{WoosleyTaam1976}
Woosley,S.E., Taam, R.E.,  1982,
ApJ 258, 761


\bibitem[Yakovlev et al. (2006)]{Yakovlev2006}
Yakovlev, D.G., Gasques, L., Wiescher M. 2006,
MNRAS, 371, 1322

\bibitem[Zdunik et al. (2008)]{Zdunik2008-AB}
Zdunik, J.L., Bejger, M., \& Haensel, P., 2008,
A\& A, 491, 489

\bibitem[Zdunik(2002)]{Zdunik2002}
Zdunik, J.L., 2002, A\& A, 394, 641
 }
\end{thebibliography}
\end{document}